# Dual Monte Carlo and Cluster Algorithms

N. Kawashima and J.E. Gubernatis
*Center for Nonlinear Studies and Theoretical Division*
*Los Alamos National Laboratory, Los Alamos, NM 87545*
(November 23, 1994)

## Abstract

We discuss the development of cluster algorithms from the viewpoint of probability theory and not from the usual viewpoint of a particular model. By using the perspective of probability theory, we detail the nature of a cluster algorithm, make explicit the assumptions embodied in all clusters of which we are aware, and define the construction of free cluster algorithms. We also illustrate these procedures by rederiving the Swendsen-Wang algorithm, presenting the details of the loop algorithm for a worldline simulation of a quantum $S = 1/2$ model, and proposing a free cluster version of the Swendsen-Wang replica method for the random Ising model. How the principle of maximum entropy might be used to aid the construction of cluster algorithms is also discussed.

pacs numbers: 02.50.+s, 02.70.+d, 05.30.-d

# I. INTRODUCTION

The development of cluster Monte Carlo algorithms by Swendsen and Wang [1] and other researchers was a significant advance in the way which computer simulations of the equilibrium properties of physical systems are implemented. These algorithms reduce the long auto-correlation times that occur as the simulations move toward a critical point. More recently, other algorithms were developed to reduce similar long times inherent to other simulations even though they are far from finite-temperature critical points [2,3]. Inspired by the work of Kandel and Domany [4], who gave a relatively general interpretation to cluster algorithms, we now propose a different perspective and also highlight several essential ingredients for developing cluster algorithms.

One of our purposes is to detail the number of natural and reasonable assumptions embodied to all cluster algorithms to-date. Our hope is to establish a framework for a more general use and more general development of cluster algorithms. Such a framework is needed. The approach of Kandel and Domany fits naturally onto classical systems defined on lattices. As such, their specifying an effective Hamiltonian and local interaction energies is quite constructive, but is unnatural if applied to quantum systems. The reason for this difference is directly traceable to the Hamiltonian in quantum mechanics being an operator and not a scalar function.

As observed by Kandel and Domany, the most effective cluster algorithms are ones in which the interactions between clusters vanish. We call these algorithms *free cluster algorithms*. Again, for quantum systems, viewing the clusters as interacting or non-interacting may be unnatural. We feel that it is best to focus on the configuration weights and directly define procedures to construct clusters that can be flipped independently. We will show that such new algorithms follow if a specific system of linear equations has a non-negative solution This system is in general underdetermined, and because underdetermined systems generally have an infinite number of solutions, if any solutions exist, finding several non-negative ones may occur. The key issue is then not the existence of a solution but rather the selection of an optimal solution, in the sense of computational efficiency. Our emphasis is defining the general structure of cluster algorithms. At this time, we only can provide standard suggestions for efficient algorithm selection.

In defining a cluster algorithm, we will argue that the standard cluster algorithm is a form of a dual Monte Carlo process. This form of Monte Carlo is a more general Monte Carlo process in which the configuration and labeling are viewed jointly. The joint probability for a configuration and a label can be expressed in terms of conditional probabilities for a label given a configuration and vice versa.

The plan of the paper is to establish notation for basic probabilities and Monte Carlo concepts in Section II. In Section III, we define what we mean by a cluster algorithm. In Section IV, we will illustrate the formalism in three different contexts: the Swendsen-Wang (SW) algorithm for a ferromagnetic Ising model, a cluster algorithm for the anisotropic quantum $S = 1/2$ quantum system, and the Swendsen-Wang replica method [2] for the random Ising model. In Section V, we conclude by discussing points for further investigation. Here, we will emphasize that they are many ways to construct cluster algorithms, but at this time we are unaware of any *a priori* way to insure one algorithm is optimal compared to others. We emphasize that our formalism is merely a procedure, as is the Kandel-Domany procedure, to start the construction of cluster algorithms. This procedure will help, but will not define how, to specify the labeling probabilities. We will discuss in the Appendix, however, how the principle of maximum entropy might be used to help accomplish this task.

## II. BACKGROUND

The common way to introduce the Monte Carlo method for simulations of the equilibrium properties starts with specifying a functional form $W(A)$ for the Boltzmann weight of states $A$ of the system and for the transition probability $T(A \to A')$ to carry $A$ to a new state $A'$. To insure the states are produced with the correct weight, the transition probabilities are almost always chosen so that the condition of detailed balance holds. This condition is

$$W(A)T(A \to A') = W(A')T(A' \to A) \qquad (2.1)$$

We will express (2.1) as

$$\Pr(A'|A)\Pr(A) = \Pr(A|A')\Pr(A') \qquad (2.2)$$

where $\Pr(A)$ is the probability of $A$ where $\Pr(A) = W(A)/Z$ with $Z = \sum_A W(A)$ being the partition function, and $\Pr(A'|A) = T(A \to A')$ is the (conditional) probability of $A'$ given $A$.

Although (2.2) seems the same as Bayes's theorem [5], its implication is different. To be more precise about the statement of detailed balance, if the Monte Carlo process produces a sequence of states $\ldots, X_{n-1}, X_n, X_{n+1}, \ldots$, then we can write the detailed balance condition as

$$\Pr(X_{n+1} = A'|X_n = A)\Pr(X_n = A) = \Pr(X_{n+1} = A|X_n = A')\Pr(X_n = A') \qquad (2.3)$$

On the other hand, Bayes's theorem is expressed as

$$\Pr(X_m = A'|X_n = A)\Pr(X_n = A) = \Pr(X_n = A|X_m = A')\Pr(X_m = A') \qquad (2.4)$$

where $n$ and $m$ are any pair of steps along the Markov chain. Bayes's theorem follows for the standard relation between joint and condition probabilities

$$\Pr(A, A') = \Pr(A'|A)\Pr(A) \qquad (2.5)$$

We will use the more detailed notation whenever we feel the distinction between the two conditions needs emphasis. The main point is however that in both cases we are dealing with probabilities. Thus, a number of relations automatically hold. The detailed balance condition is a rather special constraint imposed on the probabilities while Bayes's theorem is generally applicable to any conditional probability.

### A. Standard Monte Carlo

A Monte Carlo simulation of an equilibrium process approximates the sum over all states of a system by a sum over a smaller set of states chosen with the correct Boltzmann weight.

Each state is, in the language of probability theory, an event $A$, and the Monte Carlo process seeks to produce these events with a probability $\Pr(A)$. The Markov process in the Monte Carlo procedure is defined by the conditional probability $\Pr(A'|A)$ of state $A'$ given $A$. For this process to produce the states with probability $\Pr(A)$, several conditions must be met [6]

$$\Pr(A) \geq 0 \tag{2.6}$$

$$\sum_A \Pr(A) = 1 \tag{2.7}$$

$$\sum_{A'} \Pr(A|A') \Pr(A') = \Pr(A) \tag{2.8}$$

If $A$ can take $N$ different values, then there are $N^2$ elements for $\Pr(A|A')$. By constituting only $\mathcal{O}(N)$ constraints, the above equations illustrate the considerable freedom that exists in defining the Monte Carlo process. Typically, $N$ is a very large number so directly selecting $\Pr(A|A')$ from these equations is not practical.

Normally, a Monte Carlo process is specified so that it satisfies the detailed balance condition (2.2), which is a stronger condition than (2.8), but detailed balance still does not uniquely define $\Pr(A'|A)$. The transition probability is usually defined, in one of two ways, in terms of the ratio $\Pr(A')/\Pr(A)$. These different ways define the Metropolis and symmetric algorithms [7]. For a simple version of the symmetric algorithm

$$\Pr(A'|A) = R/(1+R), \tag{2.9}$$

where $R \equiv \Pr(A')/\Pr(A)$.

The heat-bath algorithm defines $\Pr(A'|A)$ in still another way. In the heat-bath algorithm [8], one imagines the given state $A$ being placed in contact with a heat-bath and allowed to fluctuate through various states in a manner consistent with the Boltzmann distribution $\Pr(A)$. After a while, when the heat-bath is removed, this system is left in some new state $A'$ with a probability $\Pr(A')$. The key feature is that the new state is chosen in a manner independent of the current state, i.e., $\Pr(A'|A) = \Pr(A')$. We will call a *heat-bath algorithm* those algorithms that chose the new state independently of the old state.

### B. Dual Monte Carlo

In a standard Monte Carlo algorithm, the states of the system are most naturally viewed in terms of the local variables in the Hamiltonian. These variables might simply be the values of the Ising spins at each site in a lattice, the positions of gas atoms in a box, electrons on lattice sites, etc. With these variables, a Monte Carlo process, as described above, is created by specifying the transition probability $\Pr(A'|A)$ where $A'$ is obtained from $A$, for example, by a single flip of the Ising spin at a given lattice site, the displacement of a single gas atom, etc. A number of years ago, it was recognized that the Monte Carlo process may be enhanced by introducing another set of events and performing the Markov process in a joint space [6,9]. We will adopt this modification and argue that cluster algorithms are a special case of a dual Monte Carlo process.

To develop our viewpoint, we first remark that a standard cluster algorithm starts with a state $X_0 = A$ and labels it as $Y_0 = B$ by some prescription. The label defines clusters, and the clusters are then flipped to produce a new state $A'$ that can be labeled as $B'$. The process is cycled to produce the sequence

$$\ldots \to A \to B \to A' \to B' \to A'' \to B'' \to \ldots \tag{2.10}$$

A cluster algorithm, however, can be also viewed more generally as the sequence

$$\ldots \to (A, B) \to (A', B') \to (A'', B'') \to \ldots \tag{2.11}$$

From this point of view, we would want to construct a Markov process that produces $\Pr(A, B) \equiv \lim_{n \to \infty} \Pr(X_n = A, Y_n = B)$ as its limiting probability, i.e., the transitions are viewed as from $(A, B)$ to $(A', B')$ and not just from $A$ to $A'$. Several ways exists to produce this sequence. One such way is to specify transition probabilities $\Pr(A'|A, B)$ and $\Pr(B'|A, B)$ which satisfy the following extended detailed balance condition

$$\Pr(X_{n+1} = A'|X_n = A, Y_n = B) \Pr(X_n = A, Y_n = B) =$$
$$\Pr(X_{n+1} = A|X_n = A', Y_n = B) \Pr(X_n = A', Y_n = B) \tag{2.12}$$
$$\Pr(Y_{n+1} = B'|X_{n+1} = A', Y_n = B) \Pr(X_{n+1} = A', Y_n = B) =$$
$$\Pr(Y_{n+1} = B|X_{n+1} = A, Y_n = B') \Pr(X_{n+1} = A', Y_n = B') \tag{2.13}$$

Because at equilibrium the joint probability is the same for any pair of Monte Carlo steps, we rewrite the above equations more compactly as

$$\Pr(X_{n+1} = A'|X_n = A, Y_n = B) \Pr(A, B) =$$
$$\Pr(X_{n+1} = A|X_n = A', Y_n = B) \Pr(A', B) \tag{2.14}$$
$$\Pr(Y_{n+1} = B'|X_{n+1} = A', Y_n = B) \Pr(A', B) =$$
$$\Pr(Y_{n+1} = B|X_{n+1} = A, Y_n = B') \Pr(A', B') \tag{2.15}$$

The transition probabilities $\Pr(A|A', B)$ and $\Pr(B|A, B')$ specify the algorithm. They must satisfy

$$\Pr(A, B) = \sum_{A'} \Pr(A|A', B) \Pr(A', B) = \sum_{B'} \Pr(B|A, B') \Pr(A, B') \tag{2.16}$$

In addition, we must also have

$$\Pr(A) = \sum_B \Pr(A, B) \tag{2.17}$$

to produce the desired Boltzmann weight. We also have that

$$\Pr(B) = \sum_A \Pr(A, B) \tag{2.18}$$

In a cluster algorithm, we seek to exploit the freedom we have in the choice of $\Pr(B)$ to produce an efficient and effective Monte Carlo procedure.

### C. Heat-Bath Transition Probabilities

While equations (2.14) and (2.15) express an elegant duality between the two sets of events, in a cluster algorithm these equations are used with several implicit assumptions. Most cluster algorithms of which we are aware implicitly assume that $\Pr(A'|A, B)$ is independent of $A$ and $\Pr(B'|A, B)$ is independent of $B$. We will adopt these assumptions and refer to the resulting algorithms of being of the heat-bath-type.

With the heat-bath assumptions and the use of the basic theorem of joint probabilities,

$$\Pr(A, B) = \Pr(B|A) \Pr(A) = \Pr(A|B) \Pr(B), \tag{2.19}$$

(2.14) and (2.15) reduce to

$$\Pr(X_{n+1} = A'|Y_n = B) \Pr(A|B) = \Pr(X_{n+1} = A|Y_n = B) \Pr(A'|B) \tag{2.20}$$

$$\Pr(Y_{n+1} = B'|X_{n+1} = A') \Pr(B|A') = \Pr(Y_{n+1} = B|X_{n+1} = A') \Pr(B'|A') \tag{2.21}$$

It follows that

$$\Pr(X_{n+1} = A|Y_n = B) = \Pr(A|B), \tag{2.22}$$

$$\Pr(Y_{n+1} = B|X_{n+1} = A) = \Pr(B|A), \tag{2.23}$$

which means that we should choose the transition probabilities so they agree with the limiting conditional probabilities. Therefore these equations, and hence the algorithm, depend only on the conditional probabilities $\Pr(A|B)$ and $\Pr(B|A)$. They define a dual algorithm of the heat-bath type that produces $\Pr(A)$ as the limiting distribution of the Markov chain.

If we are given $\Pr(A, B)$, the transition probabilities $\Pr(A|B)$ and $\Pr(B|A)$ are easily found. Usually, we are given $\Pr(A)$ and specify $\Pr(B|A)$. By (2.19), these two quantities are sufficient to specify $\Pr(A, B)$. Having just $\Pr(A)$ and $\Pr(A|B)$ is, in general, insufficient to fix $\Pr(A, B)$, but as we discuss below, situations exist where we can proceed in this manner and at the same time achieve considerable advantage in constructing special classes of cluster algorithms.

## III. CLUSTER ALGORITHMS

### A. Local Labeling and Other Background

In most existing cluster algorithms, the labeling of the whole system is done by labeling individual local units. In this subsection, we state this process in general terms. First, we consider systems where the given state $A$ can be represented by a set of $L$ local units described by the variable $a_i$

$$A = \{a_1, a_2, \cdots, a_L\}. \tag{3.1}$$

Each local units may consist of several local elements, each described by a variable $s_i$. Hence, we can also express the state $A$ by the set

$$A = \{s_1, s_2, \cdots, s_N\}. \tag{3.2}$$

where $L \leq N$.

For the Ising model, the local units could be bonds or plaquettes, for example. If the local units were the bonds, the local elements would be the lattice sites on the bond. For the Ising model, $s_i$ is usually a two-state (one-bit) site variable whose values are a member of the set $\{+1, -1\}$, and the values of $a_i$ are one of the four possible states that the bond $i$ may assume, i.e., the value of $a_i$ is a member of the set $\{(-1,-1),(+1,+1),(-1,+1),(+1,-1)\}$. The Ising model example also illustrates that in general a local element belongs to multiple local units, as a given site usually belongs to more than one bond.

We will restrict ourselves to systems for which we can write $W(A) = \prod_i w(a_i)$. In many respects, this condition is not very restrictive. The factorization is true for most classical Monte Carlo simulations and for some quantum Monte Carlo simulations such as those using the worldline method. In the worldline quantum Monte Carlo method for a one-dimensional system of electrons, for example, the local unit for an electron of a given electron spin is a plaquette which can have 16 different states of electron occupancy at its corners but only 6 of these states are consistent with the conservation of electron number. By considering only allowed states, one can generally express $W(A)$ in factored form.

Cluster algorithms generally assume that one can express $B$ in terms of a set of local labels $b_i$

$$B = \{b_1, b_2, \ldots, b_L\} \tag{3.3}$$

The role of these labels, assigned one to each local unit, is to define the clusters, and the values of $b_i$ by choice generally assume only a finite number of values. In the SW algorithm, the local labels of "frozen" or "deleted" are assigned to bonds. For the loop-flip algorithm for the worldline quantum Monte Carlo method, the local labels are pairs of line segments assigned the shaded plaquettes of the system. The two different segments sometimes belong to two different clusters (loops). Cluster algorithms also generally assume $\Pr(B|A) = \prod_i \Pr(b_i|a_i)$.

### B. Clustering

The essence of cluster algorithms is the changing the value of a set of many local elements, not the local units, in a coherent manner. Using our language of dual Monte Carlo, we will now define a cluster algorithm more explicitly. In what follows, we only consider the case where both the state space and the label space are discrete. The generalization to continuous variables is, for the most part, straightforward.

We start by defining a *clustering* $C$ as a function which maps the set of the serial numbers of the local elements $\mathcal{N} = \{1, 2, \ldots, N\}$ (i.e., sites, bonds, plaquettes, etc.) to a set $\mathcal{N}_c$ of the serial numbers of the clusters $\{1, 2, \cdots, N_c\}$. The integer $N_c$ is the number of clusters and is less than or equal to $N$. The value of $C(i)$ is the serial number of the cluster to which the local element $i$ belongs. The labeling process provides the basis for constructing this function.

Many ways to define clusters exist. For a collection of sites, one can create a graph by drawing a number of lines connecting pairs of sites. Those sites connected to each other, but disconnected from all other sites, form a cluster. A specific graph may contain several clusters. For a given number of sites, many different graphs may exist. Such a graphical procedure is at the heart of the Mayer cluster expansion often used in analytic studies of real gases [10].

In the SW algorithm, something more complicated is done. The connection (the label "a frozen bond") between two sites depends on the combined state of the Ising variables at those sites. Only sites mutually aligned are connected and then only with a probability chosen to insure when the cluster is flipped that the detailed balance condition is satisfied. In the SW algorithm, sites in a cluster all have the same value of the Ising spin, and the flipping process is easily defined as changing the value of spin for all sites in the cluster. For the Ising model, the original Ising spins, each of which has the value of 1 or $-1$, map to a set of cluster spins, each of which also has the value of 1 or $-1$. The beauty of the SW algorithm is that after this mapping the cluster spins do not interact. In summing over all possible configurations of clusters spins, one can place each cluster into any one of its two possible states.

In other algorithms, even more complicated things are done. In Kandel-Domany algorithm for the fully frustrated Ising model [3], the Evertz-Luna-Marcu algorithm for the six vertex model [11], and the recent algorithms for the worldline quantum Monte Carlo method [12,13], the natural local unit is a plaquette which consists of some local elements (sites), and the local labels are a set of lines connecting these sites pairwise. In these algorithms, loops replace clusters. Similar to the SW algorithm, each loop has one-bit degree of freedom, i.e., $+1$ or $-1$, which we denote by $x_i$ where $i$ specifies the loop. In contrast to the SW algorithm, not all the sites on the loop have the same state of plus or minus one, occupied or unoccupied, etc.. Therefore, the state of a site in the loop does not necessarily have the same value as the cluster variable $x_i$. Flipping the loop takes its state from one value of $x_i$ to another. What the above algorithms share with the SW algorithm is one-bit cluster variables $x_i$, the stochastic assignment of the labels, the free flipping of the clusters, and the maintenance of detailed balance.

Even more complicated situations can exist. Almost all cluster or loop algorithms to-date deal with local states that have only two possible values: spin up or spin down, occupied or unoccupied, etc. Clearly, richer physical models may have more than two values and may lead to a very rich parameterization of the state of individual clusters or loops. In this paper, we will be implicitly addressing quite general forms of clusterings. We assume that for a given state $A$ (3.1), we can assign a label $B$ (3.3), from which we can form $N_c$ clusters. Each cluster can be assigned a variable $x_i$ to describes its "state". In general, this cluster variable is not necessarily a one-bit variable. For a given cluster, the value of $x_i$ is just one from a set of values assigned to the allowed states of the cluster.

We define $\Phi(B)$ as the set of all allowed configurations consistent with the label $B$, i.e., $\Phi(B) \equiv \{A | \Pr(A, B) \neq 0\}$. In any cluster algorithm, for the global label $B$, an arbitrary state in $\Phi(B)$ is specified by a set of cluster variables

$$X = \{x_1, x_2, \ldots, x_{N_C}\}, \tag{3.4}$$

in such a way that the state of a local element depends only on the cluster variable $x_i$ where $i$ specifies the cluster to which the element belongs. In a more formal language, the essence of clustering is represented by the existence of some one-to-one mapping $f^C$ that maps the set of cluster configurations $X$ onto $\Phi(B)$. Of course, such a mapping depends on the label $B$. What we will call a *cluster Monte Carlo algorithm* is a dual Monte Carlo process defined by (2.14) and (2.15) where at least one non-trivial mapping $f^C$ exists for almost all $B$ for which $\Pr(B) > 0$. Here, a *trivial* mapping, which exists for any $\Phi(B)$, is the one in which there is only one cluster.

### C. Special cluster algorithms

In what follows, we will mainly be concerned with the question, "Can we create a cluster Monte Carlo algorithm in which the clusters can be flipped independently?" because we expect that such a cluster algorithm is advantageous both for the computational simplicity and for reducing autocorrelation times.

As stated before, a defining property of a cluster algorithm is the limited range of states to which the final state $A$ is restricted by the label $B$. We can express this situation by writing the conditional probability $\Pr(A|B)$ as

$$\Pr(A|B) = W'(A)\Delta(A,B)/N(B) \tag{3.5}$$

where the weight function $W'(A)$ is to be specified, $N(B)$ is the number of members of $\Phi(B)$, and $\Delta(A,B)$ is defined by

$$\Delta(A,B) = \begin{cases} 1, & \text{if } A \in \Phi(B) \\ 0, & \text{otherwise} \end{cases} . \tag{3.6}$$

This definition for $\Pr(A|B)$ is not unique, but it has to be chosen so that it satisfies such relations as

$$1 = \sum_A \Pr(A|B) \tag{3.7}$$

$$\Pr(A) = \sum_B \Pr(A|B)\Pr(B) \tag{3.8}$$

that are intrinsic to conditional probabilities.

In the language of the last subsection, the conditional probability $\Pr(A|B)$ means that after assigning a label $B$, we should pick a value $X$ with probability

$$\Pr(X) \equiv W'(f^C(X)) \Big/ \sum_X W'(f^C(X)). \tag{3.9}$$

Then, we map $X$ into $\Phi(B)$ with the function $f^C$ to obtain the final state $A'$. We comment that conditional probabilities cannot in general be written in the form of (3.5). In other words, when solutions exist, (3.5) defines a special class of cluster algorithms. To our knowledge, existing cluster algorithms belong to this class.

We can obtain an important additional subclass of cluster algorithms by setting $W'(A) = 1$. This selection implies that for any cluster parameterization $f^C$ of $\Phi(B)$, $\Pr(f^C(X)|B)$ is just a constant which is independent of $X$. Therefore, we can generate the final state $A$ consistent with the label $B$ by picking with equal probability a set $X$ at random, and then mapping this set into $\Phi(B)$ by the function $f^C$. Because of the clustering, the state of the system is a direct product of the states of the clusters, we pick an $X$ by picking independently and uniformly the $x_i$ from the set of possible states of cluster $i$ and collecting these values to form $X = \{x_1, x_2, \ldots, x_{N_C}\}$. Roughly speaking, the algorithms in this subclass are characterized by clusters that do not interact with each other.

As already noted, in a dual Monte Carlo algorithm of heat-bath-type, all we need to specify is $\Pr(A, B)$. To do this for the class of algorithms characterized by (3.5), we start by writing $\Pr(A) = W(A)/Z$ where $Z = \sum_A W(A)$ and then replacing $\Pr(A|B)$ in (2.17) by (3.5), we have

$$\sum_B \Delta(A, B) V(B) = W_0(A) \tag{3.10}$$

where the weight $W(A)$ is decomposed as

$$W(A) = W_0(A) W'(A) \tag{3.11}$$

and

$$V(B) = Z \Pr(B)/N(B). \tag{3.12}$$

The first factor $W_0$ is used in determining the labeling probability; the second factor $W'$, the flipping probability of clusters. Once we choose $W'(A)$ and determine $W_0(A)$ by (3.11), equation (3.10) can be viewed as a linear equation with undetermined variables $V(B)$. We emphasize that neither the uniqueness of the solution to (3.10) or its existence is guaranteed. If we obtain a solution for $V(B)$ which satisfies (3.10), we can calculate $\Pr(B|A)$ by

$$\begin{aligned} \Pr(B|A) &= \frac{\Pr(A|B) \Pr(B)}{\Pr(A)} \\ &= \frac{[W'(A) \Delta(A, B)/N(B)][V(B) N(B)/Z]}{W(A)/Z} \\ &= \frac{\Delta(A, B) V(B)}{W_0(A)}. \end{aligned} \tag{3.13}$$

In this way, we can determine both $\Pr(A|B)$ and $\Pr(B|A)$ for a given weight $W(A)$.

### D. Cluster algorithms with local labeling rules

We have reduced the task of constructing a cluster algorithm to solving (3.10) for $V(B)$, but we still have too many degrees of freedom in the algorithm to fix since in many cases the dimensionality of the label space on which $\Pr(B)$ is defined is by far greater than that of the configuration space. Therefore, to reduce further the degrees of freedom in the algorithm to make the problem tractable, we will consider a situation where $W(A)$, $W'(A)$, $\Delta(A, B)$, and $V(B)$ can be decomposed into products of local factors.

$$W(A) = \prod_i w(a_i), \tag{3.14}$$

$$W'(A) = \prod_i w'(a_i), \tag{3.15}$$

$$\Delta(A, B) = \prod_i \delta(a_i, b_i), \tag{3.16}$$

$$V(B) = \prod_i v(b_i). \tag{3.17}$$

In other words, we are considering cluster algorithms where the rules for generating a new label $B'$ for the whole system are given in terms of a collection of rules for the local elements. This situation is the case, for example, in the Swendsen-Wang algorithm for classical Potts models. Now, we can arrive at a set of equations with the number of degrees of freedom of order $\mathcal{O}(1)$

$$\sum_b \delta(a,b) v(b) = w_0(a) \tag{3.18}$$

where $w_0(a) \equiv w(a)/w'(a)$. This local equation is of central importance to this paper. Once we get a solution $v(b)$ of this equation, we can obtain the transition probability $\Pr(B|A)$ as follows

$$\Pr(B|A) = \prod_i \Pr(b_i|a_i) \tag{3.19}$$

where

$$\Pr(b|a) = \frac{\delta(a,b) v(b)}{w_0(a)}. \tag{3.20}$$

As we have seen, once a label $b$ is chosen for a local unit, the state $a$ of this unit can have only the values allowed by the matrix $\delta(a,b)$; that is, only the $a$'s for which $\delta(a,b) \neq 0$ are possible. In order that this restriction on $a$ leads to clusters in the whole system, the label $b$ must represents some clustering of the local unit, i.e., $\delta(a,b)$ has to satisfy some condition. In other words, $b$ must be such a label that *breaks up* local elements in the local unit into several groups and locks the elements in each group into a single degree of freedom.

## IV. APPLICATIONS

We now discuss several cluster algorithms from the point of view just developed. All these algorithms will be ones with local labeling rules.

### A. Swendsen-Wang algorithm for the Ising model

The Swendsen-Wang algorithm [1] is an example of a cluster algorithm with a local labeling rule which is a free cluster algorithm if the external magnetic field is zero. We will derive an algorithm for non-zero magnetic field that will reduce to the zero-field Swendsen-Wang algorithm. The Hamiltonian is

$$\mathcal{H} = -J \sum_{(i,j)} S_i S_j - H \sum_i S_i \tag{4.1}$$

with Ising variables $S_i = \pm 1$. For this model, we take the local elements to be the lattice sites, the local units to be the bonds specified by $(i,j)$, and the local variables $s_i$ to be the $S_i$. The other local variables $a_i$ have at each $i$ one of four values $(-1,-1)$, $(1,1)$, $(-1,1)$, and $(1,-1)$ which we will identify as 1, 2, 3, and 4. These values are the four allowed spin orientations on the bond.

The decomposition of the Boltzmann weight $W(A) = \exp(-\beta \mathcal{H})$ is

$$W(A) = W_0(A)W'(A) \tag{4.2}$$

where

$$W'(A) = \prod_i \exp(\beta H S_i), \tag{4.3}$$

$$W_0(A) = \prod_{(i,j)} \exp(\beta J S_i S_j) = \prod_{(i,j)} w(a_{(i,j)}). \tag{4.4}$$

Here, the local weight $w(a)$ is

$$w(1) = w(2) = r, \quad w(3) = w(4) = r^{-1}, \tag{4.5}$$

where $r \equiv \exp(\beta J)$. Thus, there are only two possible break-up operations, binding the two sites or not binding them. Accordingly, there are three possible local labels: the label $b = 1$ has $\delta(a, b) = 1$ when $a = 1$, and 2, and the label $b = 2$ has $\delta(a, b) = 1$ when $a = 3$ and 4. These two labels correspond to binding two sites. The label $b = 3$ has $\delta(a, b) = 1$ for all $a$ and corresponds to non-binding.

In matrix form, these three decompositions are

$$\delta(a, b) = \begin{pmatrix} 1 & 0 & 1 \\ 1 & 0 & 1 \\ 0 & 1 & 1 \\ 0 & 1 & 1 \end{pmatrix}. \tag{4.6}$$

Using the fact that the $w(1) = w(2)$ and $w(3) = w(4)$, we can depict this matrix as in Fig. 1. Equation (3.18) for the labeling probability reduces to

$$v(1) + v(3) = r,$$
$$v(2) + v(3) = r^{-1}.$$

and solving these equations, we obtain a set of solutions which depend on a free parameter $p$,

$$v(1) = r - p, \quad v(2) = r^{-1} - p, \quad v(3) = p, \tag{4.7}$$

where $0 \leq p \leq r^{-1}$. This last constraint is necessary to insure $v(b)$ is non-negative.

What is the best choice for $p$ is, in general, a difficult question. However, the simple guideline that *in a free cluster algorithm we should make the resulting clusters as small as possible* helps us chose a proper value. This guideline is understandable if we note that because the flipping probability of every cluster is $1/2$, the autocorrelation in the sequence of Monte Carlo data seems likely to decrease faster with a large number of small clusters than with a small number of large clusters. In the present case, this guideline suggests that we choose the largest possible $p$, i.e., $p = r$, because $v(3) = p$ is proportional to the probability of "cutting" the connection which seems necessary to promote cluster generation. Thus, our final result for the local labeling weight $v$ is

$$v(1) = 1 - r, \quad v(2) = 0, \quad v(3) = r. \tag{4.8}$$

and by using (3.20), we have

$$\Pr(b|a) = \begin{pmatrix} 1-r & 1-r & 0 & 0 \\ 0 & 0 & 0 & 0 \\ r & r & 1 & 1 \end{pmatrix}. \tag{4.9}$$

If a bond $(i, j)$ is called *satisfied* when $JS_iS_j$ is positive and *unsatisfied* otherwise, this local conditional probability tells us to cut all unsatisfied bonds and cut satisfied bonds with probability $r = e^{-2\beta J}$. This prescription is nothing but the ordinary clustering rule for the Swendsen-Wang algorithm.

Because of (3.9) and our choice of $W'$ (4.3), the flipping probability of each cluster is easily computed. The result is

$$P_{\text{flip}} = \frac{w_c^{-1}}{w_c^{-1} + w_c}. \tag{4.10}$$

Here $w_c \equiv \exp(\beta H M_c)$ where $M_c$ is the magnetization of the cluster whose absolute value equals the number of local elements in the cluster.

### B. The Swendsen-Wang replica Monte Carlo method for spin glass systems

The replica Monte Carlo method [2] was proposed by Swendsen and Wang for spin glass systems. Instead of simulating a system sequentially at different temperatures, they treat simultaneously several independent replicas of the system at different temperatures. They do this by considering a pair of systems at a time. The Hamiltonian for the pair is

$$\mathcal{H} = -\sum_{\mu=1,2} r_\mu \sum_{(i,j)} J_{i,j} S_i^{(\mu)} S_j^{(\mu)} \qquad (r_1 \leq r_2), \tag{4.11}$$

where $S_i^{(\mu)}$ is an Ising variable and $|J_{i,j}| = J$. A site is specified by two indices $(i, \mu)$. The difference $r_2 - r_1$ scales the temperature difference between the two systems.

In the replica Monte Carlo algorithm, a local element is a site $(i, \mu)$, and a local unit $[i, j]$ is a quartet of sites $(i, 1)$, $(i, 2)$, $(j, 1)$ and $(j, 2)$. For this algorithm the decomposition (3.11) is

$$W'(A) \equiv \exp(-\beta \mathcal{H}) \tag{4.12}$$
$$W_0(A) \equiv 1. \tag{4.13}$$

In other words, all the weight is in the flipping probability of clusters. The algorithm is not a free-cluster algorithm. The labeling procedure is deterministic, meaning that given $A$, $\Delta(A, B)$ is non-vanishing (i.e., unity) only for one $B$ and that

$$\Pr(B|A) = \Delta(A, B) \tag{4.14}$$
$$\Delta(A, B) = \prod_l \delta(a_l, b_l). \tag{4.15}$$

On the other hand, (3.5) reduces into

$$\Pr(A|B) = W(A)\Delta(A,B)/N(B) \tag{4.16}$$

and tells us that as long as the final state is allowed under the label $B$, the transition probability is determined by the original weight $W(A)$. From (2.9), it follows that the probability of flipping a cluster is given by

$$P = R/(1+R), \tag{4.17}$$

where $R$ is the ratio of the weights of the state $A$ before the flip and the state $A'$ after the flip, i.e., $R = W'(A')/W'(A) = W(A')/W(A)$.

The labeling rule is: If both bonds $(i,1) - (j,1)$ and $(i,2) - (j,2)$ are satisfied, or both are unsatisfied, we "freeze" the local unit. Here, *freezing* is the label under which the four sites are bound to each other. If one bond is satisfied and the other is unsatisfied, a vertical break-up is applied, which means the assignment of the label under which $(i,1)$ is bound to $(i,2)$ and $(j,1)$ is bound to $(j,2)$.

If we call "red" a pair of sites for which $S_i^{(1)} S_i^{(2)}$ is negative and call "black" any other pair, the resulting clusters are groups of red or black sites surrounded by the other color. Flipping a cluster does not change the color of the cluster, since for any $i$, the two sites $(i,1)$ and $(i,2)$ are bound to each other. Therefore, this procedure does not constitute an ergodic Markov process. It needs to be combined with an ergodic process to make the entire Markov process ergodic.

We also note that any local unit $[i,j]$ for which $i$ and $j$ belong to different clusters has one and only one satisfied bond. Thus, if we flip a cluster which includes the site $i$, this local unit contributes to the ratio $R$ in (4.17) with a factor $\exp(2\beta J(r_2 - r_1))$ or $\exp(-2\beta J(r_2 - r_1))$, according to the original orientation of spins. In particular, the flipping probability of a cluster in this algorithm becomes $1/2$ when $r_1 = r_2$. This situation illustrates why this algorithm is intended for a system with small $r_2 - r_1$: When $r_2 - r_1$ is large, the flipping probability of a large cluster is in most cases very small, and the algorithm becomes ineffective.

### C. An ergodic, free-cluster, replica method

Following our general procedure of constructing a cluster algorithm, we now will construct an ergodic, free-cluster version of the replica Monte Carlo method. For this algorithm, we choose the weight $W'(A)$ to be unity and $W_0 = \exp(-\beta \mathcal{H})$.

We consider four types of break-up operations (Fig. 2). The first one ($b=1$) is a horizontal break-up in which $(i,\mu)$ is bound to $(j,\mu)$. The second one ($b=2$) is where $(i,2)$ is bound to $(j,2)$ but $(i,1)$ and $(i,2)$ are bound to no site. The third ($b=3,4$) is the previously used vertical break-up. The last one ($b=5$) has no site bound to any other site. We note that two labels correspond to a vertical break-up; however, these two labels are equivalent because of the symmetry. Therefore, we search for a symmetric solution with respect to these two labels, that is, a solution for which $v(3) = v(4)$. For the same reason, these two labels are represented in Fig. 2 by a single column (the third column). Since the matrix $\delta(a,b)$ is the upper item of each entry in Fig. 2, the weight equation becomes

$$v(1) + v(2) + v(5) = 1,$$
$$v(2) + v(3) + v(5) = p,$$
$$v(3) + v(5) = q,$$
$$v(5) = pq, \tag{4.18}$$

where $p \equiv \exp(-2\beta r_1)$ and $q \equiv \exp(-2\beta r_2)$. The solution is

$$v(1) = (1-p)(1-q), \quad v(2) = p - q,$$
$$v(3) = v(4) = q(1-p), \quad v(5) = pq. \tag{4.19}$$

As a result, the labeling probability is given by the lower item of each entry in Fig. 2.

As we can see in Fig. 2, when $r_1 = r_2$, the label $b = 2$ is not assigned to any unit. In this case, we can argue that the resulting cluster size is generally smaller than that for the Swendsen-Wang replica algorithm: First, we note that no horizontal binding occurs in any unit where one bond is unsatisfied and the other is satisfied. This type of bonding also occurs in the Swendsen-Wang replica algorithm; however, in the present algorithm, even the vertical bonds are missing with a finite probability. Reduction in cluster size follows partially from this property. Additionally, no binding is applied to any unit where two bonds are unsatisfied while complete freezing is applied to such a unit in the Swendsen-Wang case. This second fact reduces the average cluster size even further. Still another reduction effect exists: For a unit where two bonds are satisfied, the break-up $b = 5$ is applied only with a finite probability while this type of horizontal binding occurs with probability 1 in the Swendsen-Wang case. This effect can decrease the cluster size. Therefore, the present algorithm has at least two important differences from the previous one in the case where $r_1 = r_2$: ergodicity and smaller clusters.

On the other hand, when $r_1 < r_2$, smaller clusters are not guaranteed because of the existence of the label $b = 2$ which has no counterpart in the other algorithm. This label is necessary for the present algorithm to be a free cluster algorithm. If we abandon this free-cluster property and take the difference $r_2 - r_1$ into account in the flipping probability and not in the labeling probability, then we obtain another algorithm which is similar to the Swendsen-Wang algorithm, but is one which is ergodic and produces smaller clusters. We believe, however, the free cluster version of the algorithm presented here is at least equivalent to this alternative algorithm in terms of efficiency.

### D. A Free cluster algorithm for the $S = 1/2$ $XXZ$ quantum spin model

The loop algorithm [11,14] recently proposed for the massless 6-vertex model can be applied to the quantum $S = 1/2$ problem, since in the worldline quantum Monte Carlo formulation of the problem, configurations with the quantum Boltzmann weight are equivalent to those of a special case of 6-vertex model [11]. Because the details of the application of cluster methods to quantum spin problems have not been explicitly presented elsewhere in detail, we will now give them using the language established in the present paper.

The Hamiltonian is

$$\mathcal{H} = -J \sum_{(i,j)} [\lambda(\sigma_i^x \sigma_j^x + \sigma_i^y \sigma_j^y) + \sigma_i^z \sigma_j^z], \tag{4.20}$$

where $\sigma_i^\mu$ ($\mu = x, y, z$) is a Pauli operator and the constant $\lambda$ describes the anisotropy of the problem. By using the Suzuki-Trotter decomposition formula, we can map the original problem into a problem with classical degrees of freedom in the next higher dimension. In what follows, we focus on this restatement of the problem. We will refer to the axis along the additional dimension as "vertical", and to the other axes as "horizontal;".

The local element is a site and a local unit is a shaded vertical plaquette. A site is specified by a set of indices $(i, t)$ where $i$ specifies the horizontal location and $t$ specifies the vertical location. The local variable defined on each site is a one-bit variable with the values of 0 or 1, corresponding to a down and an up z-component of spin. A vertical plaquette is one formed by four neighboring sites with two vertical edges and the other two edges being horizontal. Some fraction of vertical plaquettes are shaded. Which plaquettes are shaded depends on both the original lattice and the decomposition of the original weight by the Suzuki-Trotter formula. The shaded plaquettes are distributed across the whole lattice in such a way that no two local units share an edge (but they do share corners). Since a local unit consists of four sites, the variable defined on a unit can, in general, have 16 values. For the present problem, however, a local weight is vanishing for some of these values. We will represent a local state $a_p$ of a unit $p$ in terms of the four sites $(i, t)$, $(j, t)$, $(i, t+1)$ and $(j, t+1)$ that belong to $p$ as follows

$$a_p = \{n_{(i,t)}, n_{(j,t)}, n_{(i,t+1)}, n_{(j,t+1)}\} \tag{4.21}$$

The weight for a local unit (i.e., a shaded plaquette) is non-vanishing only when $n_{(i,t)} + n_{(j,t)} = n_{(i,t+1)} + n_{(j,t+1)}$. As a result, a local unit can have only 6 out of 16 possible states. We denote these 6 states by $1, \bar{1}, 2, \bar{2}, 3,$ and $\bar{3}$. In the site representation stated above, these states are

$$\begin{aligned} 1 &\equiv \{0,0,0,0\}, \quad \bar{1} \equiv \{1,1,1,1\}, \\ 2 &\equiv \{1,0,1,0\}, \quad \bar{2} \equiv \{0,1,0,1\}, \\ 3 &\equiv \{1,0,0,1\}, \quad \bar{3} \equiv \{0,1,1,0\}. \end{aligned} \tag{4.22}$$

Besides the constraint mentioned above, another restriction to the space of states with non-vanishing weight exists, namely, $n_{(i,t)} = n_{(i,t+1)}$ when the vertical edge $(i, t) - (i, t+1)$ does not belong to any shaded plaquette. The space $\Phi$ is defined as the set of all states $\{n_{(i,t)}\}$ that satisfy these two constraints.

The Boltzmann factor becomes

$$W(A) = \prod_p w(a_p), \tag{4.23}$$

$$\begin{aligned} w(1) &= w(\bar{1}) = \exp(\pm\tau), \\ w(2) &= w(\bar{2}) = \exp(\mp\tau)\cosh(2\lambda\tau), \\ w(3) &= w(\bar{3}) = \exp(\mp\tau)\sinh(2\lambda\tau), \end{aligned} \tag{4.24}$$

Here, $\tau$ is $\beta|J|/m$ for a Trotter number of $m$. In the case of ferromagnetic models ($J > 0$), we take the upper sign. We take the lower sign for antiferromagnetic models ($J < 0$), if the model is on non-frustrated lattices such as a square lattice.

First, we consider the ferromagnetic case. The four labels are depicted in Fig. 3(a), and the resulting weight equation is

$$v(0) + v(2) + v(3) = \exp(\tau),$$
$$v(1) + v(3) = \exp(-\tau)\cosh(2\lambda\tau),$$
$$v(1) + v(2) = \exp(-\tau)\sinh(2\lambda\tau). \quad (4.25)$$

The solution is

$$v(0) = e^\tau - p,$$
$$v(1) = (-p + e^{(2\lambda-1)\tau})/2,$$
$$v(2) = (p - e^{-(2\lambda+1)\tau})/2,$$
$$v(3) = (p + e^{-(2\lambda+1)\tau})/2, \quad (4.26)$$

where $p$ is an adjustable parameter. Here, the guideline we use to determine $p$ is the same as we used to determine $p$ in the the Swendsen-Wang algorithm for the Ising ferromagnet: we make the clusters as small as possible. In the present case, this means making $v(0)$ as small as possible. The range of $p$ is given by

$$0 \leq p \leq \min\{e^\tau, e^{(2\lambda-1)\tau}\}. \quad (4.27)$$

Therefore, in the case of $XY$-type anisotropy, i.e., $\lambda \geq 1$, we take $p = e^\tau$. The resulting labeling probability is given in Table I. From this Table, we note that the probability of assigning a break-up of type 0 is zero. In other words, the resulting clusters are simple closed loops with no branches.

On the other hand, in the case of Ising-like anisotropy, i.e., $\lambda \leq 1$, we take $p = e^{2(\lambda-1)\tau}$. Hence, the labeling probability becomes the one shown in Table II. In this case, the branching of loops is inevitable because $p(1|0) = p(\bar{1}|0) = 1 \neq 0$.

Next, we consider the antiferromagnetic case. In this case, we assign the break-up of type 0 to the states 2 and $\bar{2}$ with finite probabilities (Fig. 3(b)), in contrast to the ferromagnetic case where we assigned it to the states 1 and $\bar{1}$. The weight equation is

$$v(2) + v(3) = \exp(-\tau),$$
$$v(0) + v(1) + v(3) = \exp(\tau)\cosh(2\lambda\tau),$$
$$v(1) + v(2) = \exp(\tau)\sinh(2\lambda\tau). \quad (4.28)$$

Its solution is

$$v(0) = e^\tau - p,$$
$$v(1) = (-p + e^{(2\lambda-1)\tau})/2,$$
$$v(2) = (p - e^{-(2\lambda+1)\tau})/2,$$
$$v(3) = (p + e^{-(2\lambda+1)\tau})/2, \quad (4.29)$$

where $p$ is an adjustable parameter whose range is given by

$$0 \leq p \leq \min\{e^\tau \cosh(2\lambda\tau), e^{-\tau} + e^\tau \sinh(2\lambda\tau)\}. \quad (4.30)$$

The maximal cluster number guideline suggests that we take

$$p = e^{\tau} \cosh(2\lambda\tau) \qquad (4.31)$$

in the case of $XY$-like anisotropy and

$$p = e^{-\tau} + e^{\tau} \sinh(2\lambda\tau) \qquad (4.32)$$

in the case of Ising-like anisotropy. As a result, the labeling probabilities become the ones shown in Table III and Table IV.

We note that in both the ferromagnetic and the antiferromagnetic cases, the binding of two loops cannot be avoided for Ising-like anisotropy ($\lambda < 1$). In other words, the clusters formed in the case of Ising-like models are not simple closed loops. Roughly speaking, the clusters in this case are "bulkier" than those in $XY$-like models and spread out more in the horizontal (real-space) direction. This situation is natural, when we note that in the extremal anisotropy case, i.e., in the case of purely classical Ising models, a cluster in the Swendsen-Wang algorithm occupies a wide region of the space. We can show that this naive relation between the present algorithm and the Swendsen-Wang algorithm can be stated more clearly as follows.

When we take the classical limit, i.e., the limit of $\lambda \to 0$, the local weight factors in the ferromagnetic case are

$$\begin{aligned}
w(1) &= w(\bar{1}) \approx \exp(\tau), \\
w(2) &= w(\bar{2}) \approx \exp(-\tau), \\
w(3) &= w(\bar{3}) \approx 0.
\end{aligned} \qquad (4.33)$$

If we draw so-called worldlines by connecting the sites on which the local variables have the same value, these worldlines become straight lines because the probability of "bending" a worldline is proportional to $w(3)$ and is vanishing. A straight worldline of 1's corresponds an up-spin in the classical Ising model and a worldline of 0's corresponds a down-spin. The non-vanishing labeling probabilities are

$$p(0|1) = p(0|\bar{1}) = 1 - e^{-2\tau}, \qquad (4.34)$$
$$p(3|1) = p(3|\bar{1}) = e^{-2\tau}, \qquad (4.35)$$
$$p(3|2) = p(3|\bar{2}) = 1. \qquad (4.36)$$

Now we will consider the situation where $\tau$ is small, and hence the Suzuki-Trotter approximation is a good one and we are in the quantum limit. In such a case, $p(0|1) \approx 2\tau$ and is much smaller than $p(3|1) \approx 1 - 2\tau$. Therefore, the labeling probabilities listed above indicate that we should assign the third label to almost all the plaquettes. In other words, any one of the resulting loops is almost identical to one of worldlines, except for some that are bound to each other with a small probability. Two kinds of loops are formed: ones for which the underlying worldlines are up-spins and the ones for which the underlying worldlines are down-spins. Two adjacent loops (i.e., straight lines) of the same kind are bound to each other at the plaquette between them with a probability $1 - e^{-2\tau}$. As a result, the probability for two neighboring loops of the same kind not to be bound to each other becomes

$$(e^{-2\tau})^m = e^{-2\beta J}, \tag{4.37}$$

since there are $m$ shaded plaquettes between two adjacent straight lines.

To summarize, in the classical limit, we assign a vertical line to each site in the original representation and bind two adjacent lines of the same kind with the probability $1 - e^{-2\beta J}$. This construction is exactly the same as the ordinary Swendsen-Wang algorithm for classical Ising models if we interpret worldlines of 0's as down-spins and worldline of 1's as up-spins. By a similar argument, we can show that the $\lambda \to 0$ limit of our algorithm for the antiferromagnetic case reduces to the Swendsen-Wang algorithm for the antiferromagnetic Ising model. Thus, the present cluster algorithm is the extension of the Swendsen-Wang algorithm to quantum spin problems with $S = 1/2$. Although the critical slowing down in the quantum simulation of a $S = 1/2$ system has not extensively studied so far, obviously such a difficulty will exist. The present algorithm will be essential for reducing the difficulty in such simulations.

## V. CONCLUDING REMARKS

We have discussed the development of cluster algorithms from the viewpoint of probability theory and not from the usual viewpoint of a particular model. One of our motivations was to define general procedures for constructing clusters that are independent of the effective Hamiltonian and interaction concepts used by Kandel and Domany [4]. By using the perspective of probability theory, we clearly detailed the nature of a cluster algorithm, made explicit the assumptions embodied in all clusters of which we are aware, and defined the steps for the construction of a free cluster algorithms. We illustrated these procedures by rederiving the Swendsen-Wang algorithm, presenting the details of the loop algorithm for a worldline simulation of a quantum $S = 1/2$ model, and proposing a free cluster version of the replica method for the Ising glass.

Within the perspective of probability theory, we emphasized defining the labeling scheme embodied by the function $\delta(a, b)$ and the flipping weights $v(b)$ of the clusters, as they are actually the only things that one needs, instead of specifying the labeling probabilities, which is often done. By this shift in emphasis, we showed that the development of a cluster algorithm reduces to the solution of a linear system of equations that is generally underdetermined. A solution to this linear system is not guaranteed, but we were always able to find at least one solution by adding additional labels, if necessary. When multiple free cluster solutions exists, if we choose the one that should produce the largest number of clusters, we then typically recover existing algorithms, like the Swendsen-Wang algorithm and can also develop new free cluster algorithms, like the one for the Ising glass (Section IV B). With this approach, the development of a cluster algorithm is reduced to picking a solution of these equations (after one has specified the labeling).

We have presented the details of non-trivial examples of how one might obtain free cluster algorithms for several different systems, including a quantum mechanical problem and a spin-glass model. For these systems, the potential algorithms had a few free parameters, which were easily determined by the rule that as many clusters as possible should be formed. In other cases, they may be determined uniquely by eliminating a certain set of possible labels from consideration. For cases tested to date, we are receiving superior performance. In the Appendix, we propose a more "black-box" approach: maximizing the information theory entropy under the constraints of normalization and the linear system. For the Ising model and the Swendsen-Wang labeling, we find an algorithm similar to Swendsen-Wang's method.

A crucial condition for cluster algorithms developed so far appear to be the decomposability of $W(A)$ into a product $\prod_i w(a_i)$. As we argued, this condition is not very restrictive. All classical systems we can think of satisfy it, but some methods for simulating quantum states do not. What our formalism, or any other one, leaves unspecified is the labeling.

Specifying the labeling is the heart of the problem. Choosing labels that create clusters which approximate the large-scale coherent structures consistent with the actual physical behavior seems a natural thing to do. In general, this identification might be difficult because it requires understanding some aspects of the answer before attempting the solution. For some algorithms, like the methods for the 6-vertex model [11] and worldline method for fermions [12] and quantum spins [14], creating loops has proven to be effective and natural. The motivation for these latter algorithms was in part to maintain a local conservation condition. More recently, in a worldline method of general quantum Heisenberg spins [13], the importance of a conservation condition again appears. Loops, instead of clusters, appear. This appearance underscores once again the need for a generalized approach to cluster algorithms we presented: the presented approach is free from a specific model and concepts that might be unnatural for the problem at hand.

For several systems, creating loops has been proven to be very effective. These systems include the work of Kandel, Ben-Av, and Domany [3] on the fully frustrated Ising model, the work of Evertz, Luna, and Marcu [11] on the 6-vertex model, and our recent work on fermion and quantum spin models. We recall our discussion of the $S = 1/2$ XXZ model. In this model, for the procedures followed, whether we obtained a cluster or loop algorithm depended on the anisotropy. This result is surprising and illustrates that the goal of a cluster algorithm perhaps should not be constructing clusters or loops but rather creating whatever large-scale coherent structures that are convenient and effective.

Besides the work of Kandel and Domany, we would like to acknowledge several other works that make points related to ours. In the context of bond percolation, Ising cluster dynamics, Tamayo and Brower [15] have remarked that the cluster process can be viewed as a Monte Carlo process involving a joint probability function for the Ising variables and labels. They also pointed out that the Swendsen-Wang algorithm was not a unique way to produce a cluster algorithm for the Ising model and proceeded to develop other free cluster algorithms, one of which they demonstrated had higher efficiency. Their procedures, derived form the Kandel-Domany perspective, also lead to an underdetermined system of equations. Choosing the flipping weights, and not the labeling probabilities, is implicit in the work of Tamayo and Brower and also in the work by Evertz, Luna and Marcu on the 6-vertex model. It is explicit in the very recent work by Coddington and Han on the fully frustrated Ising model. In all these cases, the authors are lead to an underdetermined linear system of equations for which more than one acceptable solution exists. In developing the algorithms for the 6-vertex model, Evertz, Luna, and Marcu [11] used what they called "the principle of minimal freezing" to avoid the algorithm from producing unfavorable large clusters. Coddington and Han [16] also searched for solutions that avoid the production of large clusters. Thus, these workers, as ourselves, have demonstrated how to produce free cluster algorithms and possibly to avoid ones with unfavorable large clusters. We all have avoided the pitfalls of naive generalizations of the Swendsen-Wang algorithm commented upon by Domany and Kandel [4]. The unsatisfied challenge is how to find the optimal algorithm.

## ACKNOWLEDGMENTS

The work of J.E.G. was supported by the Department of Energy's High Performance Computing and Communication program at the Los Alamos National Laboratory. We thank D. Kandel for several helpful comments regarding our manuscript.

## APPENDIX

Another approach for solving (3.18), and making it more of a "black-box" procedure, is to exploit the *a priori* knowledge that $\Pr(b) = v(b)/\sum_b v(b)$ is a probability and use the principle of Maximum Entropy to assign these probabilities [5]. With this approach, the problem reduces to maximizing

$$Q = -\sum_b \Pr(b) \ln \Pr(b) + \lambda[\sum_b \Pr(b) - 1] + \sum_a \eta_a[\Pr(a) - \sum_b d(a,b)\Pr(b)] \quad (A1)$$

with respect to the $\Pr(b)$. In this equation, the $\lambda$ and $\eta_a$ are Lagrange multipliers and

$$d(a,b) = \delta(a,b)/n(b) \quad (A2)$$

where $n(b) = \sum_b \delta(a,b)$. The first term in (A1) is the information-theory entropy term. In the absence of the remaining terms, maximization would result in the probabilities $\Pr(b)$ appearing with equal weight. The second term constrains the solution to be normalized, and the third term constrains the solution to satisfy the linear equation for (3.18).

The maximization yields

$$\Pr(b) = e^{-\sum_a \eta_a d(a,b)} / \sum_b e^{-\sum_a \eta_a d(a,b)} \quad (A3)$$

and the $\eta_a$ satisfy the following set of non-linear equations:

$$\Pr(a) = \sum_b \delta(a,b) e^{-\sum_{a'} \eta_{a'} d(a',b)} / \sum_b e^{-\sum_a \eta_a d(a,b)} \quad (A4)$$

In general, these non-linear equations require numerical solution. The zero-field Ising model is simple enough that analytic solutions are possible. Using the results and definitions of Section IV A, we find for $\Pr(b)$ that

$$\Pr(1) = 1/(1 + r + r^2) \quad (A5)$$
$$\Pr(2) = r^2/(1 + r + r^2) \quad (A6)$$
$$\Pr(3) = r/(1 + r + r^2) \quad (A7)$$

where as before $r = e^{-2\beta J}$. For the Swendsen-Wang algorithm, one has from (4.8)

$$\Pr(1) = 1 - r \tag{A8}$$
$$\Pr(2) = 0 \tag{A9}$$
$$\Pr(3) = r \tag{A10}$$

The two solutions thus approach one another in the low temperature limit. We have not made numerical tests of differences in computational efficiency.

TABLES

TABLE I. The labeling probabilities for XY-like ferromagnets.

|  | b=0 | b=1 | b=2 | b=3 |
|---|---|---|---|---|
| a=1,$\bar{1}$ | 0 | 0 | $\frac{(1-e^{-2(\lambda+1)\tau})}{2}$ | $\frac{(1+e^{-2(\lambda+1)\tau})}{2}$ |
| a=2,$\bar{2}$ | 0 | $\frac{(e^{2(\lambda-1)\tau}-1)e^{2\tau}}{2\cosh(2\lambda\tau)}$ | 0 | $\frac{(e^{-2(\lambda+1)\tau}+1)e^{2\tau}}{2\cosh(2\lambda\tau)}$ |
| a=3,$\bar{3}$ | 0 | $\frac{(e^{2(\lambda-1)\tau}-1)e^{2\tau}}{2\sinh(2\lambda\tau)}$ | $\frac{(-e^{2(\lambda+1)\tau}+1)e^{2\tau}}{2\sinh(2\lambda\tau)}$ | 0 |

TABLE II. The labeling probabilities for Ising-like ferromagnets.

|  | b=0 | b=1 | b=2 | b=3 |
|---|---|---|---|---|
| a=1,$\bar{1}$ | $1-e^{-2(1-\lambda)\tau}$ | 0 | $e^{-2\tau}\sinh(2\lambda\tau)$ | $e^{-2\tau}\cosh(2\lambda\tau)$ |
| a=2,$\bar{2}$ | 0 | 0 | 0 | 1 |
| a=3,$\bar{3}$ | 0 | 0 | 1 | 0 |

TABLE III. The labeling probabilities for $XY$-like antiferromagnets.

|  | b=0 | b=1 | b=2 | b=3 |
|---|---|---|---|---|
| a=1,$\bar{1}$ | 0 | 0 | $\frac{(1-e^{2(1-\lambda)\tau})}{2}$ | $\frac{(1+e^{2(1-\lambda)\tau})}{2}$ |
| a=2,$\bar{2}$ | 0 | $\frac{e^{2(1+\lambda)\tau}-1}{2e^{2\tau}\cosh(2\lambda\tau)}$ | 0 | $\frac{1+e^{2(1-\lambda)\tau}}{2e^{2\tau}\cosh(2\lambda\tau)}$ |
| a=3,$\bar{3}$ | 0 | $\frac{e^{2(1+\lambda)\tau}-1}{2e^{2\tau}\sinh(2\lambda\tau)}$ | $\frac{1-e^{2(1-\lambda)\tau}}{2e^{2\tau}\sinh(2\lambda\tau)}$ | 0 |

TABLE IV. The labeling probabilities for Ising-like antiferromagnets.

|  | b=0 | b=1 | b=2 | b=3 |
|---|---|---|---|---|
| a=1,$\bar{1}$ | 0 | 0 | 0 | 1 |
| a=2,$\bar{2}$ | $\frac{e^{2(1-\lambda)\tau}-1}{e^{2\tau}\cosh(2\lambda\tau)}$ | $\tanh(2\lambda\tau)$ | 0 | $\frac{1}{e^{2\tau}\cosh(2\lambda\tau)}$ |
| a=3,$\bar{3}$ | 0 | 1 | 0 | 0 |

FIGURES

FIG. 1. The labels and the labeling probabilities for the Swendsen-Wang algorithm. Dashed lines in the leftmost column represent unsatisfied bonds; the solid lines, satisfied bonds. The upper item in each entry is the matrix element $\delta(a|b)$;the lower item, the labeling probability.

FIG. 2. The matrix elements $\delta(a|b)$ and the labeling probabilities for the ergodic free-cluster, replica, Monte Carlo method for spin glass systems. The upper item in each entry is the matrix element $\delta(a|b)$; the lower item, the labeling probability. The horizontal lines (dashed or solid) in each diagram in the leftmost column represent bonds at the same location but in different replicas.

FIG. 3. The labels and matrix elements $\delta(a|b)$ for the loop algorithm for (a) ferromagnetic and (b) antiferromagnetic quantum spin systems with $S = 1/2$.

| a \ b | 1 ○—○ | 2 ○—○ | 3 ○  ○ |
|---|---|---|---|
| ○—○ | 1<br>1-p | 0<br>0 | 1<br>p |
| ○----○ | 0<br>0 | 1<br>1-p/r | 1<br>p/r |

ca-fig-01 1994.03.21 N.Kawashima

| a \ b | 1 | 2 | 3,4 | 5 |
|---|---|---|---|---|
| (=, =) | 1<br>$1-a+b-ab$ | 1<br>$a-b$ | 0<br>0 | 1<br>$ab$ |
| (--, =) | 0<br>0 | 1<br>$\frac{a-b}{a}$ | 1<br>$\frac{b-ab}{a}$ | 1<br>$b$ |
| (=, --) | 0<br>0 | 0<br>0 | 1<br>$1-a$ | 1<br>$a$ |
| (--, --) | 0<br>0 | 0<br>0 | 0<br>0 | 1<br>1 |

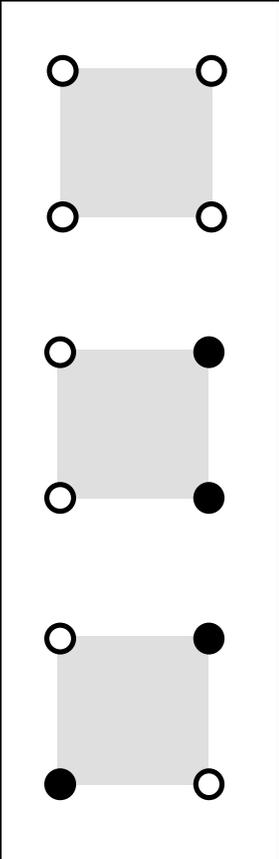

| a \ b | 0 | 1 | 2 | 3 |
|---|---|---|---|---|
| (row 1) | 0 | 1 | 0 | 1 |
| (row 2) | 1 | 1 | 1 | 1 |
| (row 3) | 0 | 0 | 1 | 1 |

`ca-fig-03b 1994.03.21 N.Kawashima`